\documentclass{article}
\usepackage[utf8]{inputenc}

\usepackage{amssymb}
\usepackage{amsmath}
\usepackage{amsthm}
\usepackage{graphicx}
\usepackage{siunitx}
\usepackage{hyperref}
\usepackage{algorithm}
\usepackage{algpseudocode}
\usepackage{xspace}
\usepackage{xcolor}

\newcommand{\bw}{\mathbf{w}}
\newcommand{\bh}{\mathbf{h}}

\newcommand{\hy}{\hat{y}}

\newcommand{\hyu}[1]{\hat{y}(u,#1)}

\newcommand{\IALS}{iALS\xspace}

\renewcommand{\O}{\mathcal{O}}

\newcommand{\myfigurewidth}{\textwidth}
\newcommand{\sourcecodelink}{https://github.com/google-research/google-research/tree/master/ials/}
\newcommand{\citeblocksolver}{\cite{rendle:blocksolver}\xspace}

\title{Revisiting the Performance of \IALS\\on Item Recommendation Benchmarks}

\author{
  \and
  Steffen Rendle\thanks{Google Research, Mountain View, USA}\\
  \texttt{srendle@google.com}
  \and
  Walid Krichene\footnotemark[1]\\
  \texttt{walidk@google.com}
  \and
  Li Zhang\footnotemark[1]\\
  \texttt{liqzhang@google.com}
  \and
  Yehuda Koren\thanks{Google, Haifa, Israel}\\
  \texttt{yehuda@google.com}
}
\date{\vspace{-6.0ex}}

\begin{document}
\maketitle

\begin{abstract} 
Matrix factorization learned by implicit alternating least squares (iALS) is a popular baseline in recommender system research publications.
iALS is known to be one of the most computationally efficient and scalable collaborative filtering methods.
However, recent studies suggest that its prediction quality is not competitive with the current state of the art, in particular autoencoders and other item-based collaborative filtering methods.
In this work, we revisit the iALS algorithm and present a bag of tricks that we found useful when applying iALS.
We revisit four well-studied benchmarks where iALS was reported to perform poorly and show that with proper tuning, iALS is highly competitive and outperforms any method on at least half of the comparisons.
We hope that these high quality results together with iALS's known scalability spark new interest in applying and further improving this decade old technique.
\end{abstract}

\section{Introduction}

Research in recommender system algorithms is largely driven by results from empirical studies.
Newly proposed recommender algorithms are typically compared to established \emph{baseline} algorithms on a set of recommender tasks.
Findings from such studies influence both the direction of future research and the algorithms for practitioners to adopt, hence it is very important to make sure that the experimental results are reliable.
Recent work has highlighted issues in recommender system evaluations (e.g.,~\cite{dacrema:reproducibility,rendle:baselines}) where newly proposed algorithms often were unable to outperform older baselines hence leading to unreliable claims about the quality of different algorithms.

In this work, we carry out a study of the performance of \IALS, a matrix factorization algorithm with quadratic loss, on top-n item recommendation benchmarks.
We demonstrate that \IALS is able to achieve much better performance than previously reported and is competitive with recently developed and often more complex models.
In addition to reinforcing the importance of tuning baseline models, we provide detailed guidance on tuning \IALS, which we hope to help unlock the power of this algorithm in practice. 

\IALS\footnote{We use the term \emph{\IALS} for \emph{implicit ALS} to refer to the method introduced in~\cite{hu:ials}. As~\cite{hu:ials} did not suggest a name for the proposed method, follow up work has used different names to refer to it. Other commonly used names are WRMF, WMF, or WALS.}~\cite{hu:ials} is an algorithm for learning a matrix factorization model for the purpose of top-n item recommendation from implicit feedback.
An example for such a recommendation task would be to find the best movies (products, songs, ...) for a user given the past movies watched (products bought, songs listened, ...) by this user.
\IALS has been proposed over a decade ago and serves as one of the most commonly used baselines in the recommender system literature.
While \IALS is regarded as a simple and computationally efficient algorithm, it is typically no longer considered a top performing method with respect to prediction quality~(e.g., \cite{liang:vae,he:ncf}). 
In this work, we revisit four well-studied item recommendation benchmarks where poor prediction quality was reported for \IALS.
The benchmarks that we pick have been proposed and studied by multiple research groups~\cite{liang:vae,he:ncf,jin:linearmodels,anelli:ncfmf,dacrema:reproducibility} and other researchers have used them for evaluating newly proposed algorithms~\cite{shenbin:recvae,kim:hvamp,lobel:ract,steck:ease}.
The poor \IALS results have been established and reproduced by multiple groups \cite{liang:vae,jin:linearmodels,he:ncf,dacrema:reproducibility,anelli:ncfmf} including a paper focused on reproducibility~\cite{anelli:ncfmf}.
However, contrary to these results, we show that \IALS can in fact generate high quality results on exactly the same benchmarks using exactly the same evaluation method.
Our \IALS numbers outperform not only the previously reported numbers for \IALS but outperform the reported quality of any other recommender algorithm on at least half of the evaluation metrics.
We attribute this surprising contradiction to the \emph{difficulty of evaluating baselines}~\cite{rendle:baselines}.
Instead of using off-the-shelf approaches for automated hyperparameter search, we achieve much better results by manual tuning.
We give a detailed description on the role of the \IALS hyperparameters, and how to tune them.

We hope that these insights help both researchers and practitioners to obtain better results for this important algorithm in the future.
As our experiments show, properly tuning an existing algorithm can have as much quality gain as inventing novel modeling techniques.
Our empirical results also call for rethinking the effectiveness of quadratic loss for ranking problems.
It is striking that \IALS achieves competitive or better performance than models learned with ranking losses (LambdaNet, WARP, softmax) which reflect the top-n recommendation task more closely.
These observations suggest that further research efforts are needed to deepen our understanding of loss functions for recommender systems.

\pagebreak

\section{Implicit Alternating Least Squares (\IALS)}

This section recaps the problem setting, model, loss and training algorithm for \IALS~\cite{hu:ials}.

\subsection{Item Recommendation from Implicit Feedback}

The \IALS algorithm targets the problem of learning an item recommender that is trained from implicit feedback.
In this problem setting, items from a set $I$ should be recommended to users $u \in U$.
For learning such a recommender, a set of positive user-item pairs $S \subseteq U \times I$ is given.
For example, a pair $(u,i) \in S$ could express that user $u$ watched movie $i$, or customer $u$ bought product $i$.
A major difficulty of learning from implicit feedback is that the pairs in $S$ are typically positive only and need to be contrasted with all the unobserved pairs $(U \times I) \setminus S$.
For example, the movies that haven't been watched by a user or the products that haven't been bought by a customer need to be considered when learning the preferences of a user.
A recommender algorithm uses $S$ to learn a scoring function $\hy : U \times I \rightarrow \mathbb{R}$ that assigns a score $\hyu{i}$ to each user-item pair $(u,i)$.
A common application of the scoring function is to return a ranked list of recommended items for a user $u$, e.g., sorting all items by $\hyu{i}$ and recommending the $k$ highest ranked ones to the user.

\subsection{Model and Loss}

\IALS uses the matrix factorization model for scoring a user-item pair.
Each user $u$ is embedded into a $d$ dimensional embedding vector $\bw_u \in \mathbb{R}^d$ and every item $i$ into a an embedding vector  $\bh_i \in \mathbb{R}^d$.
The predicted score of a user-item pair is the dot product between their embedding vectors:
Its scoring function is
\begin{align}
   \hyu{i} := \langle \bw_u, \bh_i \rangle, \quad W \in \mathbb{R}^{U \times d}, H \in \mathbb{R}^{I \times d}
\end{align}
The model parameters of matrix factorization are the embedding matrices $W$ and $H$.
These model parameters are learned by minimizing the \IALS loss.
The \IALS loss $L(W,H)$ consists of three components:
\begin{align}
    L(W,H) &= L_S(W,H) + L_I(W,H) + R(W,H) \label{eq:ials_loss} \\
    L_S(W,H) &=  \sum_{(u,i) \in S} (\hyu{i} - 1)^2 \label{eq:ials_loss_s} \\
    L_I(W,H) &= \alpha_0\sum_{u \in U} \sum_{i\in I} \hyu{i}^2 \label{eq:ials_loss_i} \\
    R(W,H) &= \lambda (\|W\|^2_F + \|H\|^2_F) = \lambda \left(\sum_{u \in U} \|\bw_u\|^2 + \sum_{i \in I} \|\bh_i\|^2\right)
\end{align}
The first component $L_S$ is defined over the observed pairs $S$ and measures how much the predicted score differs from the observed label, here $1$.
The second component $L_I$ is defined over all pairs in $U \times I$ and measures how much the predicted score differs from $0$.
The third component $R$ is an L2 regularizer that encourages small norms of the embedding vectors.
Individually, it is easy to get a loss of $0$ for each component, however, jointly they form a meaningful objective.
The trade-off between the three components is controlled by the unobserved weight $\alpha_0$ and the regularization weight $\lambda$.
Choosing the proper trade-off is crucial for \IALS and is explained in detail in Section~\ref{sec:alpha_lambda}.

\subsection{Training Algorithm}
\label{sec:algorithm}

Hu et al.~\cite{hu:ials} propose to optimize the user embeddings, $W$, and the item embeddings, $H$ by alternating minimization.
This \IALS algorithm alternates between optimizing $W$ while fixing $H$ and optimizing $H$ while fixing $W$.
Optimizing one side is equivalent to solving one linear regression problem per user if $H$ is fixed (or one problem per item if $W$ is fixed).
A key discovery of the algorithm is that the sufficient statistics from the implicit loss $L_I$ are shared between all users and can be precomputed.
With this trick, the overall computational complexity of the \IALS algorithm is $\O(d^2\,|S| + d^3\,(|U|+|I|))$.
Note that this complexity is independent of the number of all pairs $|U|\,|I|$ which makes \IALS a highly scalable algorithm with respect to the number of users and items.

A downside of the original \IALS algorithm is its quadratic/cubic runtime dependency on the embedding dimension, $d$ which becomes an issue for large embedding dimensions~\cite{pilaszy:als}.
As a solution to reduce the runtime complexity, past work has proposed to apply coordinate descent variations, where a single element of the embedding vector is optimized at a time, to lower the runtime complexity to $\O(d|S| + d^2\,(|U|+|I|))$ ~\cite{pilaszy:als,he:eals,bayer:icd}.
While this reduces the theoretical complexity, modern hardware is optimized for vector arithmetic, and scalar treatment has its drawbacks.
A compromise between the full vector treatment of the original \IALS algorithm and the scalar treatment of the follow-up work is to optimize a subvector of the embedding (a block) at a time~\citeblocksolver{}.
We use this block solver algorithm with a block size of $128$ for the experiments on ML20M and MSD; for the experiments on ML1M and Pinterest we use a standard \IALS solver.

\subsection{Frequency-based Regularization Scaling}

For rating prediction, it has been observed that vanilla stochastic gradient descent (SGD) performs much better than vanilla ALS~\cite{zhou:netflix,koren:kdd08}.
Usually, the following update rules are used for training the model parameters of matrix factorization with SGD:
\begin{align}
     \bw_u &\leftarrow \bw_u -  \eta (\langle \bw_u, \bh_i \rangle - y) \bh_i  + \lambda \bw_u)  \\
     \bh_i &\leftarrow \bh_i -  \eta (\langle \bw_u, \bh_i \rangle - y) \bw_u  + \lambda \bh_i)
\end{align}
where $\eta \in \mathbb{R}^+$ is a learning rate.
However, these SGD update rules do not optimize the loss $L_S(W,H)+R(W,H)$, i.e.,
\begin{align}
     L(W,H) = \sum_{(c,i,y) \in S} (\langle \bw_u, \bh_i \rangle - y)^2  + \lambda (\|W\|_F^2 + \|H\|_F^2) . \label{eq:sgd_mf}
\end{align}
but instead they optimize
\begin{align}
     L(W,H) &= \sum_{(c,i,y) \in S} \left[  (\langle \bw_u, \bh_i \rangle - y)^2  + \lambda \|\bw_u\|^2 +\lambda \|\bh_i\|^2\right]\\
            &= \sum_{(c,i,y) \in S} (\langle \bw_u, \bh_i \rangle - y)^2  + R^{\text{freq}}(W,H) \label{eq:sgd_loss}
\end{align}
with
\begin{align}
 R^{\text{freq}}(W,H) = \lambda \left(\sum_{u \in U} |I(u)| \|\bw_u\|^2 + \sum_{i \in I} |U(i)| \|\bh_i\|^2 \right).
\end{align}
Here, $I(u) := \{i : (u,i) \in S\}$ denotes the set of items in the training set for user $u$, and $U(i) := \{i : (u,i) \in S\}$ the set of users for item $i$.
Effectively, SGD applies heavier regularization to frequent items and users: a user $u$ gets the regularization weight $\lambda_u = \lambda |I(u)|$.
This implicit weighting of the regularizer seems to be important empirically -- as discovered by~\cite{zhou:netflix} for rating prediction.
From a probabilistic perspective, this is surprising because the regularization weight $\lambda$ corresponds to the precision of a Normal prior and typically, priors are not chosen data dependent.

We can switch between the original ALS regularizer and the frequency-based regularizer by introducing the exponent $\nu$:
\begin{align}
     R(W,H) = \lambda \left(\sum_{u \in U} |I(u)|^\nu \|\bw_u\|^2 + \sum_{i \in I} |U(i)|^\nu \|\bh_i\|^2 \right)
\end{align}
If $\nu=0$, the regularizer is equivalent to the vanilla ALS regularizer and with $\nu=1$ a frequency-based weighting as in SGD is introduced.
Values between $0$ and $1$ trade-off the effects of the two regularization schemes.

The discussion about regularization so far was motivated by rating prediction and did not contain the loss term, $L_I$, over all user-item combinations.
By adding pseudo examples with weight $\alpha_0$ for all  user-item combinations, we arrive at the following regularization weighting for \IALS:
\begin{align} \label{eq:regularization weighting}
    R(W,H) = \lambda \left(\sum_{u \in U} (|I(u)|+\alpha_0 |I|)^\nu \|\bw_u\|^2 + \sum_{i \in I}(|U(i)|+\alpha_0 |U|)^\nu \|\bh_i\|^2 \right).
\end{align}

\section{Hyperparameter Search for \IALS}

In this section, we give guidelines on how to choose hyperparameter values.
The hyperparameters of \IALS are summarized\footnote{Some \IALS implementations have a different parameterization or additional hyperparameters \cite{dacrema:reproducibility,hu:ials}.} in Table~\ref{tbl:hp}.
Performing a grid search over these six hyperparameters is likely not effective.
The poor previously reported results (see Section~\ref{sec:experiments}) that employ automated techniques indicate this.
Instead, the search should be guided by an understanding of the algorithm and the hyperparameters.
We will describe some useful metrics to plot and then some guidelines on how to explore the search space.

\begin{table*}[t]
\centering
\caption{Hyperparameters of \IALS.\label{tbl:hp}}
\begin{tabular}{|c|l|}
\hline
Symbol & Description \\
\hline
$d$ & Embedding dimension \\
$\lambda$ & Regularization \\
$\alpha_0$ & Unobserved weight \\
$T$ & Number of training iterations \\
$\sigma$ & Standard deviation for initialization of embeddings\\
$\nu$ & Frequency scaled regularization \\
\hline
\end{tabular}
\end{table*}

\subsection{Metrics}

While exploring hyperparameters, it is useful to look at several metrics.
Obviously, measuring the validation metric (e.g., Recall or NDCG on a holdout set) is useful and will drive the overall exploration.
However, the validation metrics are often only loosely connected with the training process and might not reveal why a particular hyperparameter choice does not work.
To spot some issues with learning related hyperparameters, we found it useful to plot the training loss $L$ and its components $L_S$, $L_I$, $R$ as well.
These training losses can reveal if the hyperparameters are in the wrong region and can be useful in the beginning of the hyperparameter search.

\subsection{Hyperparameters}

The metrics help to guide the hyperparameter search.
It is important to understand the hyperparameters and reduce the search space early and then focus on the important hyperparameters.
We also don't recommend tuning a large set of hyperparameters jointly but to explore them iteratively.

\paragraph{Number of Training Iterations}

It is advisable to measure the metrics during training after each iteration.
This removes the number of iterations $T$ from the search space -- provided that $T$ is large enough.
A too large value of $T$ is not a concern with respect to quality, but only with respect to runtime.
\IALS converges usually within a few iterations and a reasonable initial choice could be 16 iterations.
Depending on the observed convergence curve, this value can be increased or decreased later.
For speeding up exploration, we also found it useful to use a smaller value during initial exploration of the hyperparameter space, and then increase it for the final search.
For example, using 8 instead of 16 iterations during a broad search will cut the runtime in half.

\paragraph{Standard Deviation for Initialization}

Usually, the standard deviation for \IALS is easy to set and we haven't observed much sensitivity within a broad range.
Instead of setting the hyperparameter $\sigma$, it helps to rescale it by the embedding dimension
\begin{align}
    \sigma = \frac{1}{\sqrt{d}} \sigma^*
\end{align}
where $\sigma^*$ is a small constant, such as $0.1$.
This makes the initialization less sensitive to large changes in the embedding dimension.
The intuition is that this keeps the variance of a random dot product constant, i.e., the variance of predictions at initialization is independent of $d$.

We only observe some sensitivity if the value for the standard deviation is chosen orders of magnitude too small or too large.
In this case, it takes a few extra steps for iALS to readjust the norms of the user and item embeddings.
This can be spotted easily by plotting $L$, $L_S$, $L_I$ and $R$ where $L_S$ will not drop immediately.

\paragraph{Embedding Dimension}

The embedding dimension controls the capacity of the model.
From our experience, a common reason for suboptimal results with \IALS is that the embedding dimension is chosen too small.
We usually observe that, with proper regularization, the larger the embedding dimension the better the quality.
For example, for the Movielens 20M dataset, we found that 2000 dimensions provide the best results. 
It may seem that a 2000 dimensional embedding is too expressive for a dataset that has 73 ratings per user on average.
And even worse it might lead to overfitting.
However, empirically, larger dimensions are better and L2 regularization is very effective at preventing overfitting.
Moreover, other successful models such as VAE are also trained with large embedding dimensions and full rank models such as EASE are also effective.
The effectiveness of large embedding dimensions for matrix factorization is also well studied in the rating prediction literature~\cite{koren:advancesincf,pilaszy:als,zhou:netflix}.

Computational resources are an important factor when choosing the embedding dimension.
A good strategy is to first get a rough estimate of good hyperparameter values using a mid-sized embedding dimension, such as $d=128$, and then to perform a more refined search using larger embedding dimensions, e.g., doubling the dimension during each refinement of the other hyperparameters until the improvement plateaus.
This way, the first pass is sufficiently fast and more time can be spent for the detailed search of the most important parameters: unobserved weight and regularization.

\paragraph{Unobserved Weight and Regularization}
\label{sec:alpha_lambda}

Both unobserved weight $\alpha_0$ and the regularization $\lambda$ are crucial for \IALS\footnote{For rating prediction, the unobserved weight can be set to $\alpha_0 = 0$ and only the regularization needs to be searched.} and it is important to choose them carefully.
It is advisable to search the unobserved weight together with the regularization because both of them control the trade-off between the three loss components, $L_S$, $L_I$ and $R$.
Intuitively, we know that for item recommendation we need both $L_S$ and $L_I$ -- otherwise the solution degenerates to always predicting 1 (if $\alpha_0=0$) or always predicting 0 (if $\alpha_0 \rightarrow \infty$).
So, the observed error values of $L_S$ and $L_I$ shouldn't differ by several orders of magnitude.
Similarly, with large embedding dimensions, we need some regularization, so again the values of $R$, $L_S$ and $L_I$ should have comparable orders of magnitude. 

\begin{figure*}[t]
    \centering
    \includegraphics[width=\myfigurewidth]{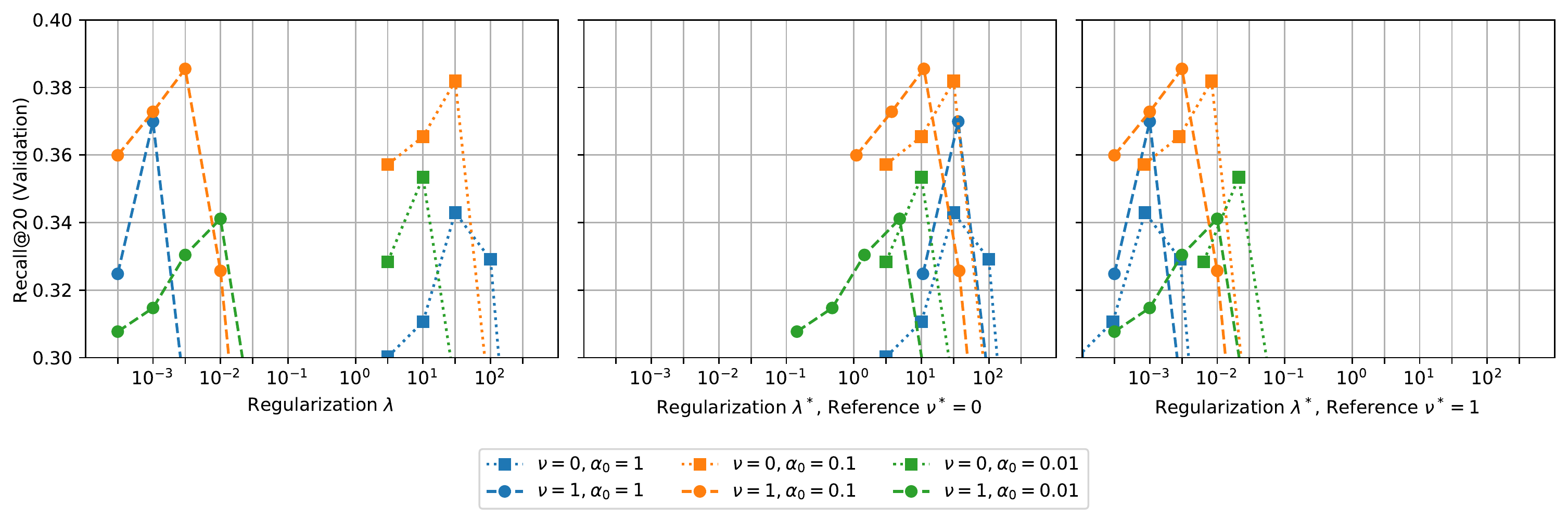}%
    \caption{When using frequency-scaled regularization, $\nu$, good regularization values $\lambda$ are in different regions for different choices of $\nu$ (left). A shifted regularization scale $\lambda^*$ brings good values on the same scale as a reference scale $\nu^*$. The middle plot shows scaling to the reference $\nu^*=0$, the right to the reference $\nu^*=1$.}
    \label{fig:scaled_reg}
\end{figure*}

The scale of regularization values depends on $\nu$~(Eq. \eqref{eq:regularization weighting}) which sets the strength of frequency regularization -- see Figure~\ref{fig:scaled_reg} left side for an example.
Without frequency regularization, $\nu=0$, good regularization values are typically larger than $1$, for frequency regularization $\nu=1$, good regularization values are usually smaller than $1$.
This is because the regularization value $\lambda$ is scaled by $(I(u) + \alpha_0 |I|)^\nu$ for each user and by $(U(i) + \alpha_0 |U|)^\nu$ for each item.
So, if $\nu$ is increased and $\lambda$ is kept constant, then the regularization effect gets stronger.
Having two parameters that interact in this way can be complicated during hyperparameter search because whenever $\nu$ is changed, the region of good values for $\lambda$ changes.
Instead it can help to normalize the regularization values to a reference scale and search over the normalized parameter $\lambda^*$ with
\begin{align}
    \lambda &= \lambda^* \frac{\sum_{i \in I} (|U(i)| + \alpha_0 |U|)^{\nu^*} + \sum_{u \in U} (|I(u)| + \alpha_0 |I|)^{\nu^*}}{\sum_{i \in I} (|U(i)| + \alpha_0 |U|)^{\nu} + \sum_{u \in U} (|I(u)| + \alpha_0 |I|)^{\nu}},
\end{align}
where $\nu^*$ is the reference scale.
For example, if we want the regularization values for all $\nu$ to be in the same region as $\nu=0$, we would choose $\nu^*=0$.
See Figure~\ref{fig:scaled_reg} middle, where good regularization values $\lambda^*$ are in the same region for $\nu=1$ as for $\nu=0$.
The right plot in Figure~\ref{fig:scaled_reg} shows a case where $\nu^*=1$ is chosen as the reference and good regularization values for $\nu=0$ are shifted to the region of the frequency regularized version $\nu=1$.
Which reference $\nu^*$ to pick depends on the practitioner and the application.
For example, if there is a comparison to SGD algorithms, choosing the reference as $\nu^*=1$ might be useful.
Or if a practitioner is more familiar with common ALS algorithms, $\nu^*=0$ might be better.
Note that the discussed rescaling of $\lambda$ does not introduce any new hyperparameters and is for convenience only.
Also $\nu^*$ is an arbitrary choice of a reference, it does not need any tuning and (unlike $\nu$) has no effects on the solution; it just simplifies the hyperparameter search for regularization values, $\lambda$.
Unless stated otherwise, in the following we discuss $\lambda^*$ with a reference point of $\nu^* = 1$.

After the relationship of the parameters has been described, we want to give some practical advice on the search.
When setting the unobserved weight one should consider the degree of matrix sparseness. It is advised that the overall magnitude of the unobserved loss term ($L_I(W,H)$~eq. \eqref{eq:ials_loss_i}) does not dominate that of the observed loss term ($L_S(W,H)$~eq. \eqref{eq:ials_loss_s}).
Thus, usually, the unobserved weight is smaller than $1.0$ and is decreasing for sparser matrices\footnote{In our loss formulation $\alpha_0=1$ is not equivalent to an unweighted SVD (PureSVD~\cite{cremonesi:topn}) because $L_I$ is defined over all pairs, not just the unobserved ones. See \cite{bayer:icd} for a discussion how the losses can be translated.}.
A good starting point is an exponential grid, for example $\alpha_0 \in \{1, 0.3, 0.1, 0.03, 0.01, 0.003\}$.
For the regularization, a good starting point is $\lambda^* \in \{0.1, 0.03, 0.01, 0.003, 0.001, 0.0003\}$.
The regularization becomes especially important for larger embedding dimensions.
If the hyperparameter search is performed on too small an embedding dimension, the regularization value found on the small dimension might not be a good one for larger dimensions.
A large enough dimension (e.g., $d=128$) can be used to get a rough estimate on which area a finer search with a higher dimension should focus on.
The suggested parameters are a starting point and need to be refined based on the results.

Some additional notes:
\begin{itemize}
    \item The training errors $L$ for different hyperparameter settings are not comparable and should not be used for selecting a model.
    \item Plotting curves for validation quality vs. log unobserved weight and validation quality vs. log regularization can help to get an idea which parts can be abandoned and where to refine. In general, we would expect to see some $\cup$-shaped curve (for error) or $\cap$-shaped curve (for quality) for the parameters -- if not, then the boundaries of the search may need to be expanded (see Figure \ref{fig:scaled_reg} for an example). Also the change in quality between two neighboring hyperparameter values shouldn't be too abrupt, otherwise more exploration is needed.
    \item It also helps not to refine exclusively around the best results, but to look at the overall behavior of the curves to get a better understanding of how hyperparameters interact on a particular data set.
    \item Measuring the noise in the validation metrics (i.e., if the same experiment is repeated twice, how much do the numbers differ) is also useful to avoid reading too much into a single experiment.
    \item At some point, large embedding dimensions should be considered. Doubling the embedding dimensions until no improvement is observed is a good strategy. If no improvement is observed, a refined search of regularization and unobserved weight might be needed. Then one can double the embedding dimension again.
\end{itemize}

\paragraph{Frequency Scaled Regularization}

In the experiments of Section~\ref{sec:experiments}, frequency scaled regularization was useful and $\nu=1$ worked the best.
Interestingly, it becomes more important with larger embedding dimensions.
For example, even though the quality plateaued with $\nu=0$, with $\nu=1$ further increasing the embedding dimension gave additional improvements.

\section{Evaluation}
\label{sec:experiments}

We revisit the performance of \IALS on four well-studied benchmarks proposed by other authors.
Two of them are item recommendation tasks and two are sampled item recommendation tasks.
We use exactly the same evaluation protocol (i.e., same splits, metrics) as in the referenced papers.
Table~\ref{tbl:data} summarizes the benchmarks.
For all quality results, we repeated the experiment 10 times and report the mean.
Our source code is available at \url{\sourcecodelink}.

\subsection{Item Recommendation}
\begin{table*}[t]
    \centering
    \caption{Benchmarks (dataset and evaluation protocol) used in our experiments. Our \IALS hyperparameters were tuned on holdout sets.
    All of the experiments share $\nu=1$ and $\sigma^*=0.1$. 
    For ML1M and Pinterest, we set a maximum dimension of $d=192$ for a fair comparison to the results from~\cite{he:ncf,rendle:ncf}.
    \label{tbl:data}}
    \setlength{\tabcolsep}{4pt}
    \begin{tabular}{|lc|rrr|SSrr|}
    \hline
        \multicolumn{2}{|c|}{Benchmark} & \multicolumn{3}{c|}{Statistics}& \multicolumn{4}{c|}{\IALS hyperparameters} \\
        Dataset       & Eval from & $|U|$  & $|I|$ & $|S|$ & {$\lambda$} & {$\alpha_0$} & T & d \\
        \hline
        ML20M~\cite{harper:movielens}              & \cite{liang:vae} & 136,677 & 20,108 & 10.0M & 0.003 & 0.1  & 16 & 2048 \\
        MSD\cite{bertinmahieux:msd}                & \cite{liang:vae} & 571,355 & 41,140 & 33.6M & 0.002 & 0.02 & 16 & 8192 \\
        ML1M~\cite{harper:movielens}       & \cite{he:ncf}    &   6,040 &  3,706 &  1.0M & 0.007 & 0.3  & 12 &  192 \\
        Pinterest~\cite{geng:pinterest}  & \cite{he:ncf}    &  55,187 &  9,916 &  1.4M & 0.02  & 0.007 & 16 & 192 \\
        \hline
    \end{tabular}
\end{table*}

In their work about variational autoencoders, Liang et al.~\cite{liang:vae} have established a set of benchmarks for item recommendation that have been followed by other authors~\cite{shenbin:recvae,kim:hvamp,steck:ease,lobel:ract} since then.
The benchmarks include results for \IALS that were produced in~\cite{liang:vae,jin:linearmodels}.
We reinvestigate the results on the Movielens 20M (ML20M) and Million Song Data (MSD) benchmarks.
We shortly recap the overall evaluation procedure and refer to~\cite{liang:vae} and their code\footnote{\url{https://github.com/dawenl/vae_cf}} for details:
The evaluation protocol removes all interactions from 10,000 users (ML20M) and 50,000 users (MSD) from the training set and puts them into a holdout set.
At evaluation time, the recommender is given 80\% of the interactions of each of the holdout users and is asked to generate recommendations for each user.
Each ranked list of recommended items is compared to the remaining 20\% of the withheld interactions, then ranking metrics are computed.
The benchmark provides two versions of the holdout data: one for validation and one for testing.
We use the validation set for hyperparameter tuning and report results for the testing set in Table~\ref{tbl:ml20m} for ML20M and Table~\ref{tbl:msd} for MSD.
The evaluation setup is geared towards algorithms that make recommendations based on a set of items, like autoencoders or item-based CF, because the evaluation users are not available during training time.
Matrix factorization with \IALS summarizes the user information in a user embedding which is usually learned at training time -- this is not possible for the evaluation users in this protocol.
We follow the evaluation protocol strictly and do not train on any part of the evaluation users.
Instead, at evaluation time, we create the user embedding using its closed form least squares expression.
As discussed in~\cite{koren:advancesincf}, matrix factorization can be seen as a item-based CF method where the history embedding (=user embedding) is generated at inference time through the closed form projection.
Note that when using a block solver~\citeblocksolver{}, the user embedding does not have a closed form expression; instead, we perform updates for each block and repeat this 8 times.

\begin{table*}
\caption{Quality results on the ML20M benchmark sorted by Recall@20 scores. \label{tbl:ml20m}}
\begin{tabular}{|l|rrr|l|}
    \hline
    Method    & Recall@20 & Recall@50 & NDCG@100  & Result from \\
    \hline
    RecVAE~\cite{shenbin:recvae}    & 0.414 & 0.553 & 0.442 & \cite{shenbin:recvae} \\
    H+Vamp (Gated)~\cite{kim:hvamp} & 0.413 & 0.551 & 0.445 & \cite{kim:hvamp} \\
    RaCT~\cite{lobel:ract}      & 0.403 & 0.543 & 0.434 & \cite{lobel:ract} \\
    Mult-VAE~\cite{liang:vae}  & 0.395 & 0.537 & 0.426 & \cite{liang:vae}\\
    LambdaNet~\cite{burges:lambdanet} & 0.395 & 0.534 & 0.427 &  \cite{shenbin:recvae} \\
    \textbf{\IALS}    & \textbf{0.395} & \textbf{0.532} & \textbf{0.425} & \textbf{our result} \\
    EASE~\cite{steck:ease}      & 0.391 & 0.521 & 0.420 &  \cite{steck:ease} \\
    CDAE~\cite{wu:cdae}      & 0.391 & 0.523 & 0.418 &  \cite{liang:vae} \\
    Mult-DAE~\cite{liang:vae}  & 0.387 & 0.524 & 0.419 &  \cite{liang:vae} \\    
    SLIM~\cite{ning:slim}      & 0.370 & 0.495 & 0.401 &  \cite{liang:vae} \\
    \textbf{\IALS} & \textbf{0.363} & \textbf{0.502} & \textbf{0.393} & \textbf{prev. result \cite{jin:linearmodels}} \\
    \textbf{\IALS}      & \textbf{0.360} & \textbf{0.498} & \textbf{0.386} &  \textbf{prev. result \cite{liang:vae}} \\
    WARP~\cite{weston:wsabie}      & 0.314 & 0.466 & 0.341 &  \cite{shenbin:recvae} \\
    Popularity& 0.162 & 0.235 & 0.191 &  \cite{steck:ease} \\
    \hline
\end{tabular}
\end{table*}

\begin{table*}
\caption{Quality results on the MSD benchmark sorted by Recall@20 scores. \label{tbl:msd}}
\begin{tabular}{|l|rrr|l|}
    \hline
    Method & Recall@20 & Recall@50 & NDCG@100  & Result from \\
    \hline
    EASE~\cite{steck:ease} & 0.333& 0.428 & 0.389 & \cite{steck:ease}\\
    \textbf{\IALS} & \textbf{0.309} & \textbf{0.415} & \textbf{0.368} & \textbf{our result} \\
    RecVAE~\cite{shenbin:recvae} & 0.276 & 0.374 & 0.326 & \cite{shenbin:recvae} \\
    RaCT~\cite{lobel:ract} & 0.268 & 0.364 & 0.319 & \cite{lobel:ract} \\
    Mult-VAE~\cite{liang:vae} & 0.266 & 0.364 & 0.316 & \cite{liang:vae} \\
    Mult-DAE~\cite{liang:vae} & 0.266 & 0.363 & 0.313 & \cite{liang:vae} \\
    LambdaNet~\cite{burges:lambdanet} & 0.259 & 0.355 & 0.308 & \cite{shenbin:recvae} \\
    \textbf{\IALS} & \textbf{0.211} & \textbf{0.312} & \textbf{0.257} & \textbf{prev. result \cite{liang:vae}} \\
    WARP~\cite{weston:wsabie} & 0.206 & 0.302 & 0.249 & \cite{shenbin:recvae} \\
    CDAE~\cite{wu:cdae} & 0.188 & 0.283 & 0.237 & \cite{liang:vae} \\
    Popularity & 0.043 & 0.068 & 0.058 & \cite{steck:ease} \\
    \hline
\end{tabular}
\end{table*}

Tables~\ref{tbl:ml20m} and \ref{tbl:msd} summarize the benchmark results.
The table highlights the previously reported numbers for \IALS and our numbers.
Previously, matrix factorization with \IALS was found to perform poorly with a considerable gap to the state of the art.
Both Mult-VAE~\cite{liang:vae} and the follow-up work RecVAE~\cite{shenbin:recvae}, H+Vamp (Gated)~\cite{kim:hvamp}, RaCT~\cite{lobel:ract} and EASE~\cite{steck:ease} outperform the previously reported \IALS results on both datasets.
However, we obtain considerably better results for \IALS matrix factorization than the previously reported \IALS numbers.
On the MSD dataset (Table~\ref{tbl:msd}) only one method, EASE, outperforms our \IALS results.
On the ML20M dataset (Table~\ref{tbl:ml20m}), \IALS has comparable performance to Mult-VAE.
No method consistently outperforms our \IALS results over both datasets and all measures, but \IALS consistently outperforms CDAE, Mult-DAE, SLIM, WARP.
For the remaining methods, \IALS is tied with 3 wins out of 6 comparisons.
It is interesting to note that most high quality results in Tables~\ref{tbl:ml20m} and \ref{tbl:msd} have been obtained by the inventors of the corresponding methods which indicates that knowledge about an algorithm is useful for obtaining good results and that automated hyperparameter search alone might not be sufficient.

\begin{figure*}
    \centering
    \includegraphics[width=\myfigurewidth]{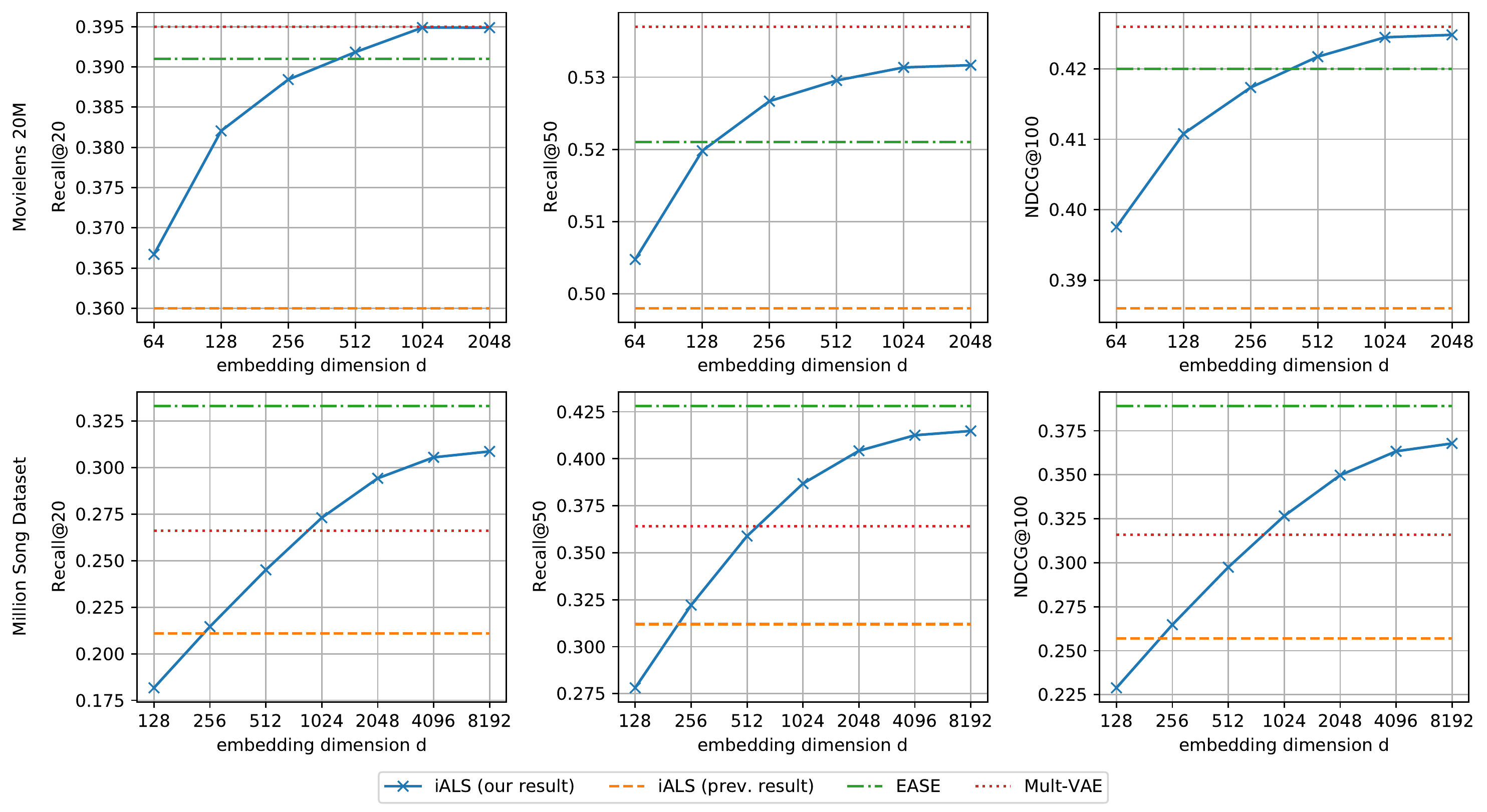}%
    \caption{Our \IALS benchmark results with a varying embedding dimension, $d$.
             For comparison, the plots contain also the previous \IALS results from~\cite{liang:vae}, EASE~\cite{steck:ease} and Mult-VAE~\cite{liang:vae}.
             The capacity for these models is not varied in the plot and the numbers represent the best values reported in previous work.
             We vary the embedding dimension for our \IALS results to get a better understanding of the importance of this hyperparameter.}
    \label{fig:ml20m_msd}
\end{figure*}

Figure~\ref{fig:ml20m_msd} breaks down the performance of \IALS by embedding dimension.
In this figure, we use the hyperparameters (see Table~\ref{tbl:data}) that were tuned for the full dimension $d=2048$ (ML20M) and $d=8192$ (MSD) and apply them to smaller dimensions.
For better comparison, we added the best numbers (independent of embedding dimension) previously reported for \IALS, EASE and Mult-VAE as reference lines. 
As can be seen, the embedding dimension has a large influence on the quality of \IALS.
Nevertheless, hyperparameter tuning is even more important: even with a small dimension ($d=64$ for ML20M and $d=256$ for MSD) the previously reported results for \IALS are already outperformed.
A dimension of $d=1024$ outperforms Mult-VAE on the MSD dataset, and a dimension of $d=512$ outperforms EASE on ML20M.
The gain of large embedding dimensions slowly levels off on ML20M around $d=1024$ and MSD around $d=8192$.
The MSD benchmark is interesting because the best performing methods, EASE and \IALS are much simpler than the advanced autoencoder methods that perform the best on ML20M.
However, both EASE and \IALS have a very large model capacity in the item representation.
EASE uses a dense item to item matrix with $|I|^2$ many parameters and for \IALS we found that a very large dimension $d=8192$ is helpful.
That might indicate that the MSD dataset has structure that is different from ML20M, for example it might have a very high rank.

Finally, the comparison so far ignored runtime.
One training epoch of a multithreaded (single-machine) C++ implementation of \IALS (using the block solver from~\citeblocksolver{}) took about 1 minute for $d=2048$ on ML20M and 40 minutes for $d=8192$ on MSD.
And for smaller dimensions: 30 seconds for $d=1024$ on ML20M and 1 minute 30 seconds for $d=1024$ on MSD.
While we ran the experiments for 16 epochs, i.e., the total runtime is 16 times higher, the results are almost converged much earlier.
For example for MSD and $d=8192$ the quality results after four epochs are Recall@20=0.305, Recall@50=0.412, NDCG@100=0.363.
It is up to the application if it is worth spending more resources to train longer.

\subsection{Sampled Item Recommendation}

We also investigate the performance of \IALS on the sampled item recommendation benchmarks from~\cite{he:ncf} which have been used in multiple publications, including~\cite{rendle:ncf,liang:vae,dacrema:reproducibility,anelli:ncfmf}.
The benchmark contains two datasets, Pinterest~\cite{geng:pinterest} and an implicit version of Movielens 1M (ML1M)~\cite{harper:movielens}.
The evaluation protocol removes one item from each user's history and holds it out for evaluation.
At evaluation time, the recommender is asked to rank a set of 101 items, where 100 of them are random items and one is the holdout item.
The evaluation metrics measure at which position the recommender places the withheld items.
As discussed in~\cite{krichene:sampled_metrics}, this evaluation task is less focused on the top items than the measured metrics indicate.
We follow the same procedure as in~\cite{rendle:ncf} for hyperparameter tuning by holding out data from training for tuning purposes.

\begin{table*}[t]
\caption{Quality results on the Movielens 1M and Pinterest benchmarks~\cite{he:ncf} sorted by HR@10 on ML1M.
         The previously reported numbers for iALS (and its eALS variation), and our \IALS numbers are in bold. The best number is underlined. \label{tbl:sampled}}
\begin{tabular}{|l|rr|rr|l|}
    \hline
     Method                                 & \multicolumn{2}{c}{ML1M}     &  \multicolumn{2}{c}{Pinterest} & Result \\
                                            & HR@10     & NDCG@10   & HR@10     & NDCG@10 & from \\
    \hline
    \textbf{\IALS}                         & \underline{\textbf{0.730}}      & \underline{\textbf{0.453}}      & \underline{\textbf{0.892}}      & \textbf{0.577}   & \textbf{our result}\\
    NeuMF~\cite{he:ncf}                                   & \underline{0.730}     & 0.447 & 0.877 & 0.552 & \cite{he:ncf} \\
    SGD-MF~\cite{koren:advancesincf}       & 0.729    & 0.452    & 0.890    & 0.579 & \cite{rendle:ncf}\\
    Mult-DAE~\cite{liang:vae}                               & 0.722     & 0.446 & 0.886 & \underline{0.580} & \cite{liang:vae} \\
    EASE~\cite{steck:ease}                               & 0.719     & 0.449 & 0.868	& 0.560 & \cite{dacrema:reproducibility} and \cite{anelli:ncfmf} \\
    SLIM~\cite{ning:slim}                    &  0.716 & 0.447 & 0.868 & 0.560    & \cite{dacrema:reproducibility} and \cite{anelli:ncfmf} \\
    \textbf{iALS}~\cite{hu:ials}           & \textbf{0.711}  & \textbf{0.438}  & \textbf{0.876} & \textbf{0.559} & \textbf{\cite{dacrema:reproducibility} and \cite{anelli:ncfmf}} \\
    \textbf{eALS}~\cite{he:eals}           & \textbf{0.704}  & \textbf{0.435} & \textbf{0.871} & \textbf{0.538} & \textbf{Fig.\,4 of \cite{he:ncf}} \\
    \hline
\end{tabular}
\end{table*}

Table~\ref{tbl:sampled} summarizes the results.
The original comparison in~\cite{he:ncf} reports results for eALS~\cite{he:eals} which is a coordinate descent variation of \IALS.
Follow up work~\cite{dacrema:reproducibility} provided results for \IALS which have been reproduced in~\cite{anelli:ncfmf}.
The previously reported performance of eALS and \IALS is poor and not competitive on both measures and both datasets.
However, with well tuned hyperparameters, we could achieve a high quality on all metrics with \IALS.
The results are very close to the ones for a well tuned SGD optimized matrix factorization~\cite{rendle:ncf} -- which is expected as both are matrix factorization models with a comparable loss.
The well-tuned \IALS is better than NCF on three metrics and tied on one.
It is also better than Mult-DAE~\cite{liang:vae} on three metrics and worse on one.
Furthermore it outperforms EASE~\cite{steck:ease} and SLIM~\cite{ning:slim} on all metrics.
Note that we limited the embedding dimension for \IALS to $d=192$ to be comparable to the previously obtained results for NCF and SGD matrix factorization.
We also produced results for smaller dimensions and $d=64$ without further hyperparameter tuning works reasonably well: for ML1M HR@10=0.722 and NDCG@10=0.445 and for Pinterest HR@10=0.892, NDCG@10=0.573.

\section{Conclusion}

This work reinvestigated matrix factorization with the \IALS algorithm and discussed techniques to obtain high quality models.
On four well-studied item recommendation benchmarks, where \IALS was reported to perform poorly, we found that it can actually achieve very competitive results when the hyperparameters are well tuned.
In particular, none of the recently proposed methods consistently outperforms \IALS.
On the contrary, \IALS outperforms any method on at least half of the metrics and datasets:
\IALS improves over a neural autoencoder in 6 out of 10 comparisons and over EASE in 7 out of 10.

These benchmarks focus on prediction quality but ignore other aspects such as training time, scalability to large item catalogs or serving of recommendations.
\IALS is known to excel in these dimensions:
(i)~It is a second order method with convergence in a few epochs.
(ii)~It uses the Gramian trick that solves the issue of $|U|\cdot|I|$ negative pairs.
(iii)~It is trivially parallelizable as it solves $|I|$ independent problems.
(iv)~It learns an embedding for each item making the model size linear in the number of items.
(v)~It uses a dot product model that allows for efficient querying of the top-n scoring items for a user.

\IALS also has some challenges:
(i)~Its loss is less aligned with ranking metrics but surprisingly, it achieves high ranking metrics on benchmarks, on par with models that optimize ranking losses such as lambdarank and softmax.
Also EASE and SLIM that share the quadratic loss with \IALS do not suffer from the quadratic loss on these benchmarks.
(ii)~In its simplest form discussed in this paper, \IALS learns a matrix factorization model, making it less flexible for richer problems with extra features.
However, \IALS has been extended for more complex models as well~\cite{hidasi:ialstensor,bayer:icd,rendle:implicit}.
(iii)~Finally, matrix factorization requires recomputing the user embedding whenever a user provides new feedback.
Bag of item models like SLIM, EASE or autoencoders do not require retraining but can just make inference using the modified input.
Nevertheless, the user embedding of \IALS has a closed form expression and this can be seen as the inference step of \IALS.
Instead of passing the history through an encoder, in \IALS a different computation (the solve step) is performed.

We hope that the encouraging benchmark results for \IALS spark new interest in this old technique.
Other models, such as autoencoders or SLIM, benefited from a growing interest in item-based collaborative filtering that resulted in improved versions such as Mult-VAE, RecVAE or EASE.
The basic \IALS model might have similar potential for improvements.
Besides academia, \IALS should be considered as a strong option for practical applications.
\IALS has very appealing properties in terms of runtime, scalability and low top-n querying costs, and as this study argues, it also performs well on benchmarks in terms of quality.

Finally, this paper is another example for the difficulty of tuning machine learning models~\cite{rendle:baselines}.
Converging to reliable numbers is a process that takes time and needs a community effort.
Over the long term, shared benchmarks like the ones established in the VAE paper~\cite{liang:vae} and adopted by other researchers~\cite{steck:ease,kim:hvamp,lobel:ract,shenbin:recvae} are a way to make progress towards reliable numbers.
Until then, it should be understood that both the benchmark results that we achieve with \IALS and the ones from the other methods might be further improved in the future.

\bibliographystyle{acm}

\begin{thebibliography}{10}

\bibitem{anelli:ncfmf}
{\sc Anelli, V.~W., Bellog\'{\i}n, A., Di~Noia, T., and Pomo, C.}
\newblock Reenvisioning the comparison between neural collaborative filtering
  and matrix factorization.
\newblock In {\em Fifteenth ACM Conference on Recommender Systems\/} (New York,
  NY, USA, 2021), RecSys '21, Association for Computing Machinery,
  p.~521–529.

\bibitem{bayer:icd}
{\sc Bayer, I., He, X., Kanagal, B., and Rendle, S.}
\newblock A generic coordinate descent framework for learning from implicit
  feedback.
\newblock In {\em Proceedings of the 26th International Conference on World
  Wide Web\/} (2017), WWW '17, pp.~1341--1350.

\bibitem{bertinmahieux:msd}
{\sc Bertin-mahieux, T., Ellis, D. P.~W., Whitman, B., and Lamere, P.}
\newblock The million song dataset.
\newblock In {\em In Proceedings of the 12th International Conference on Music
  Information Retrieval (ISMIR\/} (2011).

\bibitem{burges:lambdanet}
{\sc Burges, C., Ragno, R., and Le, Q.}
\newblock Learning to rank with nonsmooth cost functions.
\newblock In {\em Advances in Neural Information Processing Systems\/} (2007),
  B.~Sch\"{o}lkopf, J.~Platt, and T.~Hoffman, Eds., vol.~19, MIT Press.

\bibitem{cremonesi:topn}
{\sc Cremonesi, P., Koren, Y., and Turrin, R.}
\newblock Performance of recommender algorithms on top-n recommendation tasks.
\newblock In {\em Proceedings of the Fourth ACM Conference on Recommender
  Systems\/} (New York, NY, USA, 2010), RecSys '10, Association for Computing
  Machinery, p.~39–46.

\bibitem{dacrema:reproducibility}
{\sc Dacrema, M.~F., Boglio, S., Cremonesi, P., and Jannach, D.}
\newblock A troubling analysis of reproducibility and progress in recommender
  systems research.
\newblock {\em ACM Trans. Inf. Syst. 39}, 2 (Jan. 2021).

\bibitem{geng:pinterest}
{\sc {Geng}, X., {Zhang}, H., {Bian}, J., and {Chua}, T.}
\newblock Learning image and user features for recommendation in social
  networks.
\newblock In {\em 2015 IEEE International Conference on Computer Vision
  (ICCV)\/} (2015), pp.~4274--4282.

\bibitem{harper:movielens}
{\sc Harper, F.~M., and Konstan, J.~A.}
\newblock The movielens datasets: History and context.
\newblock {\em ACM Trans. Interact. Intell. Syst. 5}, 4 (Dec. 2015),
  19:1--19:19.

\bibitem{he:ncf}
{\sc He, X., Liao, L., Zhang, H., Nie, L., Hu, X., and Chua, T.-S.}
\newblock Neural collaborative filtering.
\newblock In {\em Proceedings of the 26th International Conference on World
  Wide Web\/} (Republic and Canton of Geneva, Switzerland, 2017), WWW '17,
  International World Wide Web Conferences Steering Committee, pp.~173--182.

\bibitem{he:eals}
{\sc He, X., Zhang, H., Kan, M.-Y., and Chua, T.-S.}
\newblock Fast matrix factorization for online recommendation with implicit
  feedback.
\newblock In {\em Proceedings of the 39th International ACM SIGIR Conference on
  Research and Development in Information Retrieval\/} (New York, NY, USA,
  2016), SIGIR '16, Association for Computing Machinery, p.~549–558.

\bibitem{hidasi:ialstensor}
{\sc Hidasi, B., and Tikk, D.}
\newblock Fast {ALS}-based tensor factorization for context-aware
  recommendation from implicit feedback.
\newblock In {\em Joint {European} {Conference} on {Machine} {Learning} and
  {Knowledge} {Discovery} in {Databases}\/} (2012), Springer, pp.~67--82.

\bibitem{hu:ials}
{\sc Hu, Y., Koren, Y., and Volinsky, C.}
\newblock Collaborative filtering for implicit feedback datasets.
\newblock In {\em Proceedings of the 2008 Eighth IEEE International Conference
  on Data Mining\/} (2008), ICDM '08, pp.~263--272.

\bibitem{jin:linearmodels}
{\sc Jin, R., Li, D., Gao, J., Liu, Z., Chen, L., and Zhou, Y.}
\newblock Towards a better understanding of linear models for recommendation.
\newblock In {\em Proceedings of the 27th ACM SIGKDD Conference on Knowledge
  Discovery \& Data Mining\/} (New York, NY, USA, 2021), KDD '21, Association
  for Computing Machinery, p.~776–785.

\bibitem{kim:hvamp}
{\sc Kim, D., and Suh, B.}
\newblock Enhancing vaes for collaborative filtering: Flexible priors \& gating
  mechanisms.
\newblock In {\em Proceedings of the 13th ACM Conference on Recommender
  Systems\/} (New York, NY, USA, 2019), RecSys '19, Association for Computing
  Machinery, p.~403–407.

\bibitem{koren:kdd08}
{\sc Koren, Y.}
\newblock Factorization meets the neighborhood: A multifaceted collaborative
  filtering model.
\newblock In {\em Proceedings of the 14th ACM SIGKDD International Conference
  on Knowledge Discovery and Data Mining\/} (New York, NY, USA, 2008), KDD '08,
  ACM, pp.~426--434.

\bibitem{koren:advancesincf}
{\sc Koren, Y., and Bell, R.}
\newblock {\em Advances in Collaborative Filtering}.
\newblock Springer US, Boston, MA, 2011, pp.~145--186.

\bibitem{krichene:sampled_metrics}
{\sc Krichene, W., and Rendle, S.}
\newblock On sampled metrics for item recommendation.
\newblock In {\em Proceedings of the 26th ACM SIGKDD International Conference
  on Knowledge Discovery \& Data Mining\/} (New York, NY, USA, 2020), KDD '20,
  Association for Computing Machinery, p.~1748–1757.

\bibitem{liang:vae}
{\sc Liang, D., Krishnan, R.~G., Hoffman, M.~D., and Jebara, T.}
\newblock Variational autoencoders for collaborative filtering.
\newblock In {\em Proceedings of the 2018 World Wide Web Conference\/}
  (Republic and Canton of Geneva, CHE, 2018), WWW '18, International World Wide
  Web Conferences Steering Committee, p.~689–698.

\bibitem{lobel:ract}
{\sc Lobel, S., Li, C., Gao, J., and Carin, L.}
\newblock Ract: Toward amortized ranking-critical training for collaborative
  filtering.
\newblock In {\em 8th International Conference on Learning Representations,
  {ICLR} 2020, Addis Ababa, Ethiopia, April 26-30, 2020\/} (2020).

\bibitem{ning:slim}
{\sc Ning, X., and Karypis, G.}
\newblock Slim: Sparse linear methods for top-n recommender systems.
\newblock In {\em Proceedings of the 2011 IEEE 11th International Conference on
  Data Mining\/} (USA, 2011), ICDM '11, IEEE Computer Society, pp.~497--506.

\bibitem{pilaszy:als}
{\sc Pil\'{a}szy, I., Zibriczky, D., and Tikk, D.}
\newblock Fast als-based matrix factorization for explicit and implicit
  feedback datasets.
\newblock In {\em Proceedings of the Fourth ACM Conference on Recommender
  Systems\/} (New York, NY, USA, 2010), RecSys '10, Association for Computing
  Machinery, p.~71–78.

\bibitem{rendle:implicit}
{\sc Rendle, S.}
\newblock Item recommendation from implicit feedback.
\newblock {\em CoRR abs/2101.08769\/} (2021).

\bibitem{rendle:ncf}
{\sc Rendle, S., Krichene, W., Zhang, L., and Anderson, J.}
\newblock Neural collaborative filtering vs. matrix factorization revisited.
\newblock In {\em Proceedings of the 14th ACM Conference on Recommender
  Systems\/} (2020), RecSys '20.

\bibitem{rendle:blocksolver}
{\sc Rendle, S., Krichene, W., Zhang, L., and Koren, Y.}
\newblock {iALS++}: Speeding up matrix factorization with subspace
  optimization, 2021.

\bibitem{rendle:baselines}
{\sc Rendle, S., Zhang, L., and Koren, Y.}
\newblock On the difficulty of evaluating baselines: {A} study on recommender
  systems.
\newblock {\em CoRR abs/1905.01395\/} (2019).

\bibitem{shenbin:recvae}
{\sc Shenbin, I., Alekseev, A., Tutubalina, E., Malykh, V., and Nikolenko,
  S.~I.}
\newblock Recvae: A new variational autoencoder for top-n recommendations with
  implicit feedback.
\newblock In {\em Proceedings of the 13th International Conference on Web
  Search and Data Mining\/} (New York, NY, USA, 2020), WSDM '20, Association
  for Computing Machinery, p.~528–536.

\bibitem{steck:ease}
{\sc Steck, H.}
\newblock Embarrassingly shallow autoencoders for sparse data.
\newblock In {\em The World Wide Web Conference\/} (New York, NY, USA, 2019),
  WWW '19, Association for Computing Machinery, p.~3251–3257.

\bibitem{weston:wsabie}
{\sc Weston, J., Bengio, S., and Usunier, N.}
\newblock Wsabie: Scaling up to large vocabulary image annotation.
\newblock In {\em Proceedings of the Twenty-Second International Joint
  Conference on Artificial Intelligence - Volume Volume Three\/} (2011),
  IJCAI'11, AAAI Press, pp.~2764--2770.

\bibitem{wu:cdae}
{\sc Wu, Y., DuBois, C., Zheng, A.~X., and Ester, M.}
\newblock Collaborative denoising auto-encoders for top-n recommender systems.
\newblock In {\em Proceedings of the Ninth ACM International Conference on Web
  Search and Data Mining\/} (New York, NY, USA, 2016), WSDM '16, Association
  for Computing Machinery, p.~153–162.

\bibitem{zhou:netflix}
{\sc Zhou, Y., Wilkinson, D., Schreiber, R., and Pan, R.}
\newblock Large-scale parallel collaborative filtering for the netflix prize.
\newblock In {\em Proceedings of the 4th International Conference on
  Algorithmic Aspects in Information and Management\/} (Berlin, Heidelberg,
  2008), AAIM '08, Springer-Verlag, pp.~337--348.

\end{thebibliography}

\end{document}